\documentstyle[12pt,aps,prc,epsfig,preprint]{revtex}
\tightenlines
\begin{document}
\begin{titlepage}

\title{\bf Antiproton-nucleus potentials from global fits to
antiprotonic X-rays and radiochemical data}

\author{E.~Friedman $^a$, A.~Gal $^a$, J.~Mare\v{s} $^b$ \\
$^a${\it Racah Institute of Physics, The Hebrew University,
Jerusalem 91904, Israel} \\
$^b${\it Nuclear Physics Institute, 25068 \v{R}e\v{z}, Czech Republic}
}

\maketitle
\begin{abstract}

We report on global fits of optical-model parameters to 90 data points 
for $\bar p$ X-rays and 17 data points of radiochemical data put together. 
By doing separate fits to the two kinds of data it is possible 
to determine phenomenologically the radial region where the absorption 
of antiprotons takes place and to obtain neutron densities which represent 
the average behaviour over the periodic table. A finite-range attractive 
and absorptive $\bar p$-nuclear isoscalar potential fits the data well. 
Self-consistent dynamical calculations within the RMF model demonstrate 
that the polarization of the nucleus by the {\it atomic} antiproton 
is negligible. 

\vspace{7mm} 

$PACS$ 13.75.Cs  14.20.Dh  21.10.Gv  36.10.-k

\vspace{7mm}

\newline 
Corresponding author: E. Friedman, elifried@vms.huji.ac.il
\newline
tel: +972 2 658 4667, fax: +972 2 658 6347
\end{abstract}

\vspace{7mm}

\centerline{\today}
\end{titlepage}

\section{Introduction}
\label{sec:int}

Experiments on strong-interaction effects in antiprotonic atoms
usually involve measurements of X-rays from $\bar p$-atomic transitions,
providing values of level shifts and widths, the latter either measured
directly or indirectly via the observed yields. Global analyses of
large data sets, in terms of an optical model approach, have provided
parameters for $\bar p$-nucleus optical potentials \cite{BFG95,BFG97}
which fit well all the data then available, close to 50 data points, 
across the periodic table. It was found that the radial region which 
is `sampled' by the $\bar p$-atom data is approximately between the 
half-density nuclear radius and some 4 fm outside of it. The data 
could be well fit by a strongly attractive and strongly absorptive 
isoscalar $\bar p$-nuclear potential $V_{\rm opt}$, underlined by 
a ${\bar N}N$ $s$-wave interaction, in which the imaginary (absorptive) 
part outweighs the real (attractive) part. Gradient ($p$-wave) terms 
in $V_{\rm opt}$ were not required by the $\bar p$-atomic data. 
Furthermore it was concluded, for any {\it global} fit of 
$\bar p$-atom data, that the isovector components of $V_{\rm opt}$ 
were relatively insignificant, provided that single-particle densities 
which describe correctly the nuclear surface region and beyond were used.
We note that since $V_{\rm opt}$ is strongly absorptive, its extrapolation 
into nuclear-matter densities is highly model dependent. 
Even if a unique extrapolation were possible, the existence of well-defined 
{\it nuclear} bound states of antiprotons would have been unlikely due 
to the large widths expected. 

The experimental situation has changed significantly with the availability 
of high-quality data for several sequences of isotopes along the periodic 
table due to the PS209 collaboration \cite{TJC01} which made it possible 
to perform global fits to larger and more accurate data bases consisting 
of close to 100 data points. A good description of these new data was 
achieved in global analyses by using a finite-range $V_{\rm opt}$ in which 
a ${\bar p}N$ $s$-wave interaction is folded with the nuclear density 
$\rho = \rho_n + \rho_p$, as reported preliminarily in Ref. \cite{FGa04}. 
Another type of experimental 
information on the $\bar p$-nucleus interaction in $\bar p$ atoms is 
obtained using the so-called radiochemical method. By studying the 
production of nuclei differing from the target nucleus by the removal 
of one neutron or one proton it is possible to obtain information on 
the ratios of $\bar p$ absorption on neutrons to $\bar p$ absorption 
on protons in the periphery of the target nucleus, at a distance 
about 2.5 fm beyond the half-density charge radius according to 
Ref. \cite{WSS96}. At such distances the nuclear density is dominated by 
the neutron density which is not directly measured or model-independently 
determined, and it is vital to use reliable nuclear models rooted in 
a sound nuclear phenomenology to extract these densities for use in 
optical-model calculations. Close to 20 target nuclei have been studied 
by the radiochemical method \cite{LJT98,SHK99} and in several cases 
\cite{TJL01,STC03,KWT04} good consistency was found between the 
radiochemical data and the X-ray data, within an optical-model approach. 

The present paper reports on global fits of optical potentials
to all those experimental results, consisting of 90 points of X-ray 
data {\it and} 17 points of radiochemical data put together. Our aim 
in this paper is to determine the $\bar p$-nuclear interaction potential 
at the nuclear surface and beyond, while identifying simultaneously 
the largely unknown neutron densities which provide a good fit to the 
combined data in terms of $V_{\rm opt}$ that depends sensitively on 
these densities. The organization of the paper is as follows. 
In section \ref{sec:rmf} we verify that the conventional approach of 
{\it static} atomic calculations, i.e. using static nuclear densities, 
is valid. This is achieved by performing {\it dynamical} calculations 
within a relativistic mean field (RMF) approach where the nucleus is 
allowed to be polarized by the atomic $\bar p$. Note that for 
$\bar p$-nuclear configurations, in contrast, dynamical RMF calculations 
for antiprotons \cite{MSB05} have yielded considerable nuclear 
polarization effects. Section \ref{sec:pot} presents the $\bar p$-nucleus 
potential and the methodology of the present work, in particular as 
regarding the use of neutron densities. 
In section \ref{sec:res} we present results, first only for the X-ray
data, followed by results for the radiochemical data where we examine
under what conditions the two sets of results are consistent with each 
other. Global fits to the combined data base of X-ray and radiochemical 
data are then presented and conclusions are drawn on the admissible 
neutron-density shapes. Section \ref{sec:con} summarizes the findings 
of this work.

\section{Dynamical calculations of antiprotonic atoms}
\label{sec:rmf}

Self-consistent RMF calculations of deeply bound $\bar p$ - {\it nuclear} 
states \cite{MSB05}, as well as $\bar K$-nuclear states \cite{MFG05}, 
revealed extremely large dynamical polarization effects. The core nucleus 
is highly compressed and its binding energy significantly increases due to
the strongly attractive $\bar p$ - nucleus interaction. 
It is therefore mandatory to check to what extent polarization effects
are important in the calculations of $\bar p$-{\it atomic} states and 
whether the static approach hitherto used in atomic calculations is 
justified. In order to address this problem, we performed self-consistent 
dynamical calculations of $\bar p$ - atomic states within the RMF model.

In the RMF framework the core nucleons and the  $\bar p$ interact through 
the exchange of isoscalar-scalar ($\sigma$) and -vector ($\omega$), 
isovector-vector ($\rho$), and Coulomb ($A$) fields, which are treated 
in the mean-field approximation. The advantage of the RMF approach is that 
antinucleons ($\bar N$) are naturally included in the Lagrangian density 
and, consequently, the derivation of the relevant equations of motion is
straightforward. The Dirac equation for nucleons and antinucleons 
($i=N$, $\bar N$) reads:
\begin{equation}
\label{equ:Dir}
\left[ -{\rm i}{\vec\alpha}{\cdot}{\vec\nabla} + 
\beta(M_i + S_i)+V_i\right]{\Psi_i}^{a} =
\epsilon_i^{a} {\Psi_i}^{a} \; ,
\end{equation}
where $S_i = g_{\sigma i}\sigma$, $V_i = g_{\omega i}\omega^0 + g_{\rho i}
\tau_3{\rho^0}_{\!\!\! 3} + e_i \frac{(1+\tau_3)}{2} A^0$, 
and $a$ denotes quantum numbers of single-particle states.

The presence of the $\bar p$ in the nuclear system under consideration 
modifies the source terms in the equations of motion for the meson fields:

\begin{eqnarray}
\label{equ:mes}
\left(-\Delta + m_{\sigma}^2 + c_1\sigma +c_2\sigma^2 \right) \sigma \; &=&
- g_{\sigma N}\rho_{SN}- g_{\sigma {\bar N}}\rho_{S{\bar N}} \; ,
\nonumber \\
\left(-\Delta + m_{\omega}^2 + d {\omega_0}^2 \right) \omega_0 \; &=&
g_{\omega N}\rho_{BN} + g_{\omega {\bar N}}\rho_{B{\bar N}} \; ,\\
\left(-\Delta + m_{\rho}^2  \right) {\rho^0}_{\!\!\! 3} \; &=&
g_{\rho N}\rho_{3N}+ g_{\rho {\bar N}}\rho_{3{\bar N}} \; ,\nonumber \\
\left(-\Delta \right) A_0 \; &=&
e_{p}\rho_{p}+ e_{\bar p}\rho_{\bar p} \; ,\nonumber
\end{eqnarray}
where $\rho_{Si}$, $\rho_{Bi}$, $\rho_{3i}$, $\rho_{j}$ ($j=p$, $\bar p$) 
are the scalar, vector (baryon), isovector, and proton and antiproton 
densities, respectively.

While for nucleons the Dirac equation (\ref{equ:Dir}) was solved, 
for the antiproton we actually solved a Schr{\" o}dinger equation 
with a complex $\bar p$ optical potential. The real part of the 
$\bar p$ potential, ${\rm Re}V_{\rm opt}$, was constructed as the 
Schr{\"o}dinger equivalent potential from the scalar and vector 
meson mean fields:
\begin{equation} 
\label{equ:Vpbar} 
{\rm Re}V_{\rm opt}=S_{\bar N}+V_{\bar N}+(S_{\bar N}^2 -V_{\bar N}^2)/2M~~  .
\end{equation} 
Note that, for antiprotons, both $S_{\bar N}$ {\it and} $V_{\bar N}$ are 
attractive owing to G-parity. 
When solving Eqs.(\ref{equ:mes}) for the meson mean fields, we make the
approximation $\rho _{S \bar N}=\rho _{B \bar N}=\rho _{\bar N}$,
where $\rho _{\bar N}$ is the $\bar p$ density formed by solving
the Schr\"odinger equation (\ref{equ:Vpbar}). This approximation amounts
to neglecting a few percent difference at most between the scalar and 
vector densities, a neglect that turns out to be completely insignificant
for the dynamical effects calculated in $\bar p$ atoms.
The $\bar p$ optical potential is dominated 
by its imaginary term \cite{BFG97} which is due to the annihilation of 
antiprotons on nucleons. Im$V_{\rm opt}$ was taken proportional to the 
nuclear density $\rho (r)$ and its depth was adjusted to fit the 
atomic data. It is to be noted that in this dynamical approach the nuclear 
density $\rho$ is not a static quantity but is affected by the antiproton
interacting strongly with the core nucleons via meson mean fields. 

In our dynamical calculations of the $\bar p$-nucleus systems,
the standard RMF Lagrangian with the nonlinear parameterizations
TM1 \cite{STo94} and due to Sharma et al. \cite{SNR93} were used.
Applying the G-parity transformation as a starting point to determine the 
antinucleon ($\bar p$) couplings $g_{j \bar N}$ ($j=\sigma,\omega,\rho$) 
but realizing that the presence of strong annihilation channels and 
various many-body effects could cause significant deviations from the
G-parity values, we introduced following Ref.\cite{MSB05} a reduction 
parameter $\varepsilon$ ($\varepsilon \leq 1$) to allow for departures 
from the ideal G-parity limit: 
\begin{equation} 
\label{equ:gparity}
g_{\sigma \bar N} = \varepsilon g_{\sigma N}, \;\;
g_{\omega \bar N} = - \varepsilon g_{\omega N}\, , \;\;
g_{\rho N} = \varepsilon g_{\rho N}\, .
\end{equation}
The reduction parameter $\varepsilon$ was determined by fitting the
calculated energy shift and width
of a particular  $\bar p$ atomic state to the corresponding
experimental result. Static calculations showed that fits to the atomic
$\bar p$ data for $^{16}$O, $^{40}$Ca and $^{208}$Pb produced very nearly 
the same potential parameters as obtained from fits to the full data base 
(but with larger errors with which the optical-potential parameters are 
determined). 
Consequently we chose these three nuclei for checking  polarization effects 
by studying the $3d$ atomic state of $^{16}$O, the $4f$ state of $^{40}$Ca 
and the $9k$ state of $^{208}$Pb.
The value of the reduction parameter $\varepsilon$
needed to fit the atomic data
 was between 0.15 ($^{40}$Ca) and 0.35 ($^{208}$Pb).
This indicates a large
deviation of the $\bar p$ couplings from the G-parity values.

The coupled system of equations (\ref{equ:Dir}) for nucleons, 
(\ref{equ:mes}) for the meson- and electromagnetic fields, and the 
Schr{\"o}dinger equation for $\bar p$ was solved self-consistently
by iterations. In order to check the numerical procedure, 
$\bar p$-{\it nuclear} states were calculated using the TM1 parameterization 
and the results of Ref.\cite{MSB05} were reproduced. Large polarization 
effects were found for $\bar p$-nuclear states, as expected.
Subsequently, dynamical effects for $\bar p$-{\it atomic} states
were studied. Comparing the energy shifts and widths calculated 
self-consistently with the predictions of the corresponding static 
calculations, we found totally negligible effects of less than 1 eV 
for the above three $\bar p$-atomic states. 

Finally, deeply bound $\bar p$-atomic states \cite{FGa99} were also 
calculated in $^{16}$O, $^{40}$Ca, and $^{208}$Pb.
For illustration, Table \ref{tab:tab1} presents calculated binding 
energies ($B_{\bar p}$) and widths ($\Gamma_{\bar p}$) for the $1s$
and $2p$ $\bar p$-atomic states in Ca, comparing between the results 
of dynamical and of static calculations. The differences of few eV, 
caused by the polarization effects, are exceedingly small and 
experimentally unobservable. We conclude that static analyses of 
$\bar p$-atomic data which neglect the rearrangement of the core 
nucleons are adequate.

\section{Methodology}
\label{sec:pot}

In order to preserve the connection to previous studies of hadronic atoms 
\cite{BFG97}, the Klein-Gordon (KG) equation is used in the form 
\begin{equation}
\label{equ:KG}
\left[\Delta - 2{\mu}(B+V_{\rm opt}+V_c) + (V_c+B)^2 \right]{\psi} = 0~~ ~~
(\hbar = c = 1).
\end{equation} 
Here, $V_c$ denotes the static Coulomb potential for the $\bar p$ due to 
the finite charge distribution of the nucleus, including the first-order 
$\alpha (Z\alpha)$ vacuum-polarization potential, 
$\mu$ is the $\bar p$-nucleus reduced mass and
$B=B_{\bar p}+{\rm i}{\Gamma_{\bar p}}/2$ is the complex binding energy.  
The interaction of antiprotons with the nucleus is described here in terms
of an optical potential $V_{\rm opt}$ which in the simplest  
`$t \rho$' form is given by
\begin{equation}
\label{equ:potl}
2\mu V_{{\rm opt}}(r) = -4\pi(1+\frac{\mu}{M}
\frac{A-1}{A})[b_0(\rho_n+\rho_p)
  +b_1(\rho_n-\rho_p)]~~,
\end{equation}
where $\rho_n$ and $\rho_p$ are the neutron and proton density
distributions normalized to the number of neutrons $N$ and number
of protons $Z$, respectively, $A=N+Z$, and $M$ is the mass of the 
nucleon. In the zero-range `$t \rho$' approach the parameters 
$b_0$ and $b_1$ are minus the $\bar p$-nucleon isoscalar 
and isovector scattering lengths, respectively, otherwise these parameters
may be regarded as `effective' and are obtained from fits to the data, 
either as given above or in a finite-range folding model \cite{FGa04}. 
The explicit inclusion of an isovector term in $V_{\rm opt}$ 
is motivated by the fact that much data now exist \cite{TJC01} 
for chains of isotopes and because the radiochemical method provides 
information on ratios of neutron to proton densities. 
A least-squares search procedure was
used to adjust the potential parameters so as to obtain a best fit
to the experimental shift and width measurements.

A note on wave equations is in order here. 
The quadratic term $(V_c+B)^2$ in the KG equation (\ref{equ:KG})
has little effect on the calculated strong-interaction shift
and width, and omitting it gives rise to the nonrelativistic
Schr\"odinger equation. One might ask why use the KG equation as extension
into the relativistic domain instead of the Dirac equation for the 
$\bar p$ fermion.
Indeed when interpreting experimental
transition energies in order to extract the strong interaction effects
it is essential to use the Dirac equation with finite size nuclear
charge distribution and vacuum polarization terms \cite{SHE98}.
However,  strong interaction effects in $\bar p$ atoms are normally
given as  the proper averages over the fine structure components.
The use of the KG equation rather than the Dirac equation  
 is numerically justified when fine-structure effects are negligible 
or are treated in an average way, as for the X-ray transitions considered here. 
The leading $j$ dependence ($j = l \pm \frac{1}{2}$) of the energy for 
solutions of the Dirac equation for a point-charge $1/r$ potential 
goes as $(j + \frac{1}{2})^{-1}$, and on averaging it over 
the projections of $j$ gives rise to $(l + \frac{1}{2})^{-1}$ which is 
precisely the leading $l$ dependence of the energy for solutions of the 
KG equation. The higher-order contributions to the spin-orbit splitting 
are suppressed by $O(Z \alpha /n)^2$ which is of order 1$\%$ for the 
high-$n$ X-ray transitions encountered for antiprotons. 
We checked numerically 
for few typical cases that the spin-orbit averaged shifts and widths 
thus obtained differ by less than 1\% from the $(2j+1)$-average of the 
corresponding quantities obtained by solving the Dirac equation. This 
difference is considerably smaller than the experimental 
errors placed on the measured X-ray transition energies and widths. 

The nuclear densities are an essential ingredient of the optical potential. 
The density distribution of the protons is usually considered known as
it is obtained from the nuclear  charge distribution \cite{FBH95} by
unfolding the finite size of the charge of the proton. 
The neutron distributions are, however, generally not known to sufficient 
accuracy. A host of different methods have been
applied to the extraction of rms radii of neutron distributions in nuclei
but the results are sometimes conflicting, see Refs. 
\cite{BFG89,GNO92,SHi94,KFA99,CKH02,JTL04}. 
For many nuclei there is no direct experimental information whatsoever 
on neutron densities and one must then rely on models. 
To complicate things further we note that there is a long history of
conflict between values of neutron rms radii derived from
experiments using hadronic projectiles and neutron rms radii obtained
from theoretical calculations. For that reason we have adopted a
semi-phenomenological approach that covers a broad range of possible
neutron density distributions.

Experience with pionic atoms showed \cite{FGa03} that
the feature of  neutron density distributions which is most relevant
in  determining strong interaction effects
in pionic atoms is the radial extent, as represented for
example by $r_n$, the neutron density rms radius. Other features
such as  the detailed shape of the distribution have only minor effect. 
For that reason we chose the rms radius as the prime parameter in the 
present study. Since $r_p$, the rms radius for the proton density 
distribution, is considered to be known, we focus attention on values 
of the difference $r_n-r_p$. In previous analyses of $\bar p$ 
radiochemical data \cite{TJL01,JTL04} a linear dependence of $r_n-r_p$ 
on $(N-Z)/A$ was employed, namely, 
\begin{equation} \label{equ:RMF}
r_n-r_p = \alpha \frac{N-Z}{A} + \gamma \; ,
\end{equation}
with $\alpha$ close to 1.0 fm and $\gamma$ close to zero. The same expression
with $\alpha$ close to 1.5 fm was found \cite{FGa03} to represent well
results of RMF calculations \cite{LRR99} for stable nuclei, but these 
values of $r_n-r_p$ are larger by about $0.05 - 0.10$ fm than the 
`experimental' values in medium-weight and heavy nuclei used in 
recent relativistic Hartree-Bogoliubov (RHB) versions of mean-field 
calculations \cite{NVF02,LNV05}. Expression (\ref{equ:RMF}) has been 
adopted in the present work and, for lack of better global information
about neutron densities, the value of $\alpha$ was varied over a reasonable
range in fitting to the data. This procedure is based on the expectation 
that for a large data set over the whole of the periodic table some local 
variations will cancel out and that an average behaviour may be established.

In order to allow for possible differences in the shape of the neutron
distribution, the `skin' and the `halo' forms of Ref. \cite{TJL01} were
used, as well as an average between the two. Adopting a two-parameter
Fermi distribution both for the proton (unfolded from the charge distribution)
and for the neutron density distributions

\begin{equation} 
\label{equ:2pF} 
\rho_{n,p}(r)  = \frac{\rho_{0n,0p}}{1+{\rm exp}((r-R_{n,p})/a_{n,p})} \; ,
\end{equation}
then for each value of $r_n-r_p$ in the `skin' form
the same diffuseness parameter for the protons and the neutrons
$a_n=a_p$ was used and the $R_n$ parameter was determined from the
rms radius $r_n$.
In the `halo' form
the same radius parameter $R_n=R_p$ was assumed and $a_n^{\rm h}$ was
determined from $r_n$. In the `average'
option the diffuseness parameter was set to be the average of the
above two diffuseness parameters $a_n^{{\rm ave}}=(a_p+a_n^{\rm h})/2$
and the radius parameter $R_n$ was then determined from the
rms radius $r_n$. In this
way we have used three shapes of the neutron distribution for each value
of its rms radius all along the periodic table.
Figure \ref{fig0} shows as an example the densities for $^{120}$Sn
after a finite range folding (see below). The relevant radial region
for the $\bar p$-nucleus interaction is 8-9 fm in this case, and it is
seen that the neutron densities are quite different for the three models.

Returning to the optical potential Eq.(\ref{equ:potl}), we note that in
Ref.\cite{FGa04} it was found that the X-ray data could be described better
with a finite-range interaction, where the densities of Eq.(\ref{equ:potl})
were replaced by `folded' densities
\begin{equation}
\label{equ:fold}
\rho ^F (r)~~=~~\int d{\bf r}' \rho({\bf r}') \frac{1}{\pi ^{3/2} \beta^3}
e^{-({\bf r}-{\bf r'})^2/\beta^2}~~,
\end{equation}
with the value of $\beta$ = 0.85 fm used for both the real and imaginary
parts of the potential.
For the analysis of the radiochemical data we adopt the approach of
Refs.\cite{LJT98,SHK99,TJL01} that the method is sensitive to the
neutron to proton density ratio close to 2.5 fm outside of the
half-density radius of the charge density \cite{WSS96}.  The experimental
ratios of absorption on neutrons to absorption on protons were therefore
compared to
\begin{equation}
\label{equ:ratio}
\frac{{\rm Im}(b_0+b_1)I_n}{{\rm Im}(b_0-b_1)I_p}
\end{equation}
where $I_{n,p}$ are the volume integrals of the neutron and proton densities,
respectively, either between 2.0  and 3.0 fm or between 2.5 and 3.5 fm
outside of the half-density radius of the charge density. For a finite-range
potential the folded densities were used.
No use was made of atomic wavefunctions in calculating the ratios because 
their effect  largely cancel out in the ratios. Choosing the range 
of integration 
was guided by the conclusions of Ref. \cite{WSS96} which obviously were
based on atomic wavefunctions. By requiring consistency between radiochemical
data and X-ray data we could independently check the conclusions of
Ref. \cite{WSS96} as explained below.

\section{Results and Discussion}
\label{sec:res}

\subsection{X-ray data}
\label{subsec:X}
A brief report on global fits to the present extended X-ray data has
already been published \cite{FGa04}. The data base included strong
interaction level shifts and widths for antiprotonic atom levels
in $^{16,18}$O, $^{40,42,43,44,48}$Ca, $^{54,56,57,58}$Fe,
$^{58,60,62,64}$Ni, $^{90,96}$Zr, $^{106,116}$Cd, $^{112,116,120,124}$Sn,
$^{122,124,126,128,130}$Te and $^{208}$Pb, 
a total of 90 points \cite{TJC01}. 
It was found that by introducing a finite range into the interaction
within the folding-model approach of Eq.({\ref{equ:fold}) consistently
better fits to the data were obtained compared to the corresponding
zero-range approach.

 Figure \ref{fig1} shows values of $\chi ^2$ obtained
for the three shapes of neutron distributions discussed above as function
of the slope parameter $\alpha$ in the expression for $r_n-r_p$,
Eq.(\ref{equ:RMF}), using a finite-range parameter of $\beta$=0.85 fm.
It is seen that both the `halo' model and the `average' model for $\rho _n$
produce very acceptable fits with $\chi ^2$ per point of about 2.2.
Recall that values of $\alpha$ are likely to be between 0.9 and
1.3 fm, as the value of 1.5 fm found from fits to results of RMF calculations
is now regarded as an over-estimate. The results for the
`halo' version confirm
earlier findings \cite{TJL01} that for this shape of neutron density
agreement with experiment is obtained for $\alpha$ very close to 1 fm.
For the `skin' variety of $\rho _n$ the minimum of $\chi ^2$ is obtained
for a value of $\alpha$  close to 2 fm which is
unreasonably large \cite{JTL04,NVF02,LNV05}. The `average' type of
$\rho _n$ is quite acceptable both in the quality
of fit and in the value of $\alpha$ at the minimum of $\chi ^2$.
For this model the slope of 
$\rho _n$ is smaller at large radii than the slope of $\rho _p$, 
as expected on grounds of binding energies, yet
the model is not restricted in having the same half-density radii for
the proton and for the neutron distributions. Note that from global fits
to strong interaction shifts and widths in pionic atoms it is
found \cite{FGa03} that the `halo' shape is rejected, whereas the other 
two shapes are quite acceptable. However, the pionic atom data are 
sensitive to considerably smaller radii (and consequently larger 
densities) than is the case for $\bar p$ atoms.
Figure \ref{fig2} shows values of the isoscalar amplitude $b_0$ from
fits to the data for the `average' type of neutron density using a
finite-range interaction. The isovector amplitude was found to be
consistent with zero \cite{FGa04} and we return to this point below.

\subsection{Radiochemical data}
Experimental ratios of $\bar p$ absorption on neutrons to absorption on
protons were taken from Refs. \cite{LJT98,SHK99}. Initial calculations
showed that very large contributions to the resulting $\chi ^2$ came
from $^{106}$Cd and $^{112}$Sn and subsequently these two nuclei were
excluded from the data set. Possible explanations for the problem
with these two nuclei in terms of a $\bar p p$ quasi bound state are given
in Ref. \cite{Wyc01}.
We have therefore used 17 values of absorption ratios
for the following nuclei: $^{48}$Ca, $^{58}$Ni, $^{96}$Zr,
$^{100}$Mo, $^{96,104}$Ru, $^{116}$Cd, $^{124}$Sn, $^{128,130}$Te,
$^{144,154}$Sm, $^{148}$Nd, $^{160}$Gd, $^{176}$Yb, $^{232}$Th and
$^{238}$U. With the potential parameter $b_1$ consistent with zero
the ratios Eq.(\ref{equ:ratio}) become independent of the parameters 
of the potential, but they are quite sensitive to values of $r_n-r_p$.
This is seen clearly in Fig. \ref{fig3} where $\chi ^2$ values for the
radiochemical data are shown as function of $\alpha$ for the `halo'
and for the `average' types of neutron densities.
With the proton density held fixed the calculated ratios  obviously
depend on the neutron density {\it and} on the radius where the ratios
of densities are calculated. To avoid the use of densities at a point 
we have integrated the densities over a distance of 1 fm, and 
Fig. \ref{fig3} shows results for integrations between 2.5 and 3.5 fm 
and between 2 and 3 fm beyond the half-density radius of the charge 
distribution. However, due to the exponential decrease of the densities 
at such large radii the integrals are dominated by the densities close 
to the lower limit of the range of integration.

Comparing positions of the minima in Fig. \ref{fig3} for the 
radiochemical data with 
Fig. \ref{fig1} for the X-ray data, we note  a consistency between 
the two regarding the parameter $\alpha$ when the integration range 
for the radiochemical data is 2.5 to 3.5 fm beyond the charge radius, 
whereas moving that range down by 0.5 fm leads to inconsistency between 
the two types of data. This is a significant test because both types 
of data depend differently on the nuclear densities. This result 
confirms in a phenomenological way the theoretical conclusion of 
Wycech et al. \cite{WSS96} that most of the absorption takes place 
close to 2.5 fm beyond the charge radius. We therefore adopt the 2.5 
to 3.5 fm segment as the integration range for the global analysis of 
the combined X-ray  and the radiochemical data.

\subsection{Global fits to X-ray and radiochemical data}
\label{subsec:com}

Fits to all the 107 data points due to the X-ray and radiochemical data
were made, using the various shapes for the neutron densities as
described above and varying the isoscalar potential parameter $b_0$
and the isovector parameter $b_1$. Finite-range folding was assumed
throughout with a range parameter $\beta$=0.85 fm.  For the radiochemical data
the integration range was chosen between 2.5 and 3.5 fm beyond the
charge half-density radius, as described above. The real part of $b_1$ was
always found to be consistent with zero and subsequently it was excluded
from the fits. Figure \ref{fig4} shows results for the `average' shape
of the neutron densities and Fig. \ref{fig5} shows similar results
for the `halo' shape. In the lower part of the figures we show values
of $\chi ^2$ for the X-ray data separately and for the combined X-ray and
radiochemical data. The upper part shows the potential parameters.

It is seen that although Im$b_1$ assumes non-vanishing values
away from the minima of $\chi ^2$, 
 at the minima this parameter is essentially
zero. Accepting those minima for the `halo' and for the `average' shapes
with $\alpha$ between 0.9 and 1.3 fm as a fair representation,
{\it on the average}, of neutron densities, we end up with the values
of $b_0$ as found in Ref.\cite{FGa04}, namely Re$b_0$=1.3$\pm$0.1 fm,
Im$b_0$=1.9$\pm$0.1 fm, with a folding range of $\beta$=0.85 fm.

The vanishing of Im$b_1$ at the best-fit points for the two acceptable models
for neutron densities deserves a comment. We note that the four $\bar N N$
potentials considered in Table 3 of 
Ref. \cite{BFG95} yield for Im$b_1$ values which
are a factor of 10-20 smaller than the corresponding values of Im$b_0$,
so it is conceivable that also for the {\it effective} phenomenological
parameters a similar situation will hold. With the many groups of
isotopes in the present data base the dependence on $N-Z$ must play
a role. The neutron densities are obviously properly normalized, but
the rms radii are not set {\it a priori} and they are found, on the
average, in the process of $\chi ^2$ fits. The vanishing of Im$b_1$
is presumably an indication of the general validity of the neutron densities
used here. As a test we note that for the unacceptable `skin' model,
where  $\alpha$ is close to 2 fm at the minimum, a non-vanishing
value is found for Im$b_1$, which presumably compensates for the
unphysical neutron densities.

The fits shown in Figs. \ref{fig1} - \ref{fig5} can be marginally 
improved, without changing the conclusions of the present work, if the 
parameterization of $r_n - r_p$ in Eq. (\ref{equ:RMF}) is slightly modified
to incorporate the expected $A$ dependence of the parameter $\gamma$ 
due to the increased Coulomb repulsion within the $N=Z$ core. 
To this end we have used $\gamma = -0.0162~A^{1/3}$ fm to interpolate 
between $r_n - r_p$ values in $^{16}$O, $^{40}$Ca and the Sn 
and Pb isotopes within recent RHB calculations \cite{NVF02} which yield 
values for the $(N-Z)/A$ slope parameter $\alpha \sim 1.5 - 1.6$ fm. 
This analysis, again, strongly suggests that the prefered neutron densities 
are best described by the `average' type shape. A more expanded 
discussion of these results, plus related ones for pionic atoms, 
are relegated to a separate publication.

\section{Conclusions}
\label{sec:con}

In conclusion, self-consistent dynamical calculations within 
the RMF model demonstrate that the polarization of the nucleus 
by the atomic antiproton is negligible, thus confirming the 
validity of the common static approach in which the nuclear 
polarization is disregarded. 
We have performed, for the first time, global fits of
optical model parameters to 90 data points for antiprotonic X-rays
and 17 data points of radiochemical data put together. With the
help of separate fits to the two kinds of data we could verify
in a phenomenological way the theoretical prediction \cite{WSS96} 
for the radial region where most of the absorption of atomic 
antiprotons takes place. Despite the use of many groups of isotopes 
all along the periodic table, the parameters of the isovector part 
of the potential could not be determined in {\it global} fits, 
and are therefore assumed to be consistent with zero. 
The optical potentials are determined at radii where the density
is 2-3\% of the central nuclear density. It is therefore rather
meaningless to extrapolate these potentials to the nuclear interior.
However, if extrapolated with the nuclear densities used in the
present work, then at the center of the nucleus the potential is attractive
with about 110 MeV depth, and its absorptive part is close 
to 160 MeV deep. While this attraction is fairly strong, about twice that 
for nucleons in nuclear-matter densities, it is much weaker than naive 
G-parity arguments suggest. 
The world's data on $\bar p$-nucleus interaction potential are
well accounted for with the effective parameters Re$b_0$=1.3$\pm$0.1 fm, 
Im$b_0$=1.9$\pm$0.1 fm, with a folding range of $\beta$=0.85 fm.

\vspace{7mm}

This work was supported in part by the Israel Science Foundation
grant 131/01 and  by the GA AVCR grant IAA1048305.

\begin{table}
\caption{Binding energies $B_{\bar p}$ and widths $\Gamma_{\bar p}$ (in keV) 
of deeply-bound $\bar p$ atomic states in $^{40}$Ca calculated within the 
static and dynamical approaches using the TM1 RMF parameterization.}
\label{tab:tab1}
\begin{center}
\begin{tabular}{ccccc}
$nl$ & static $B_{\bar p}$ & dynamical $B_{\bar p}$ & 
static $\Gamma_{\bar p}$ & dynamical $\Gamma_{\bar p}$  \\
\hline
 $1s$ & 1835.132 & 1835.129 & 584.970 & 584.967 \\

 $2p$ & 1516.105 & 1516.101 & 353.299 & 353.300 \\
\end{tabular}
\end{center}
\end{table}

\begin{figure}
\epsfig{file=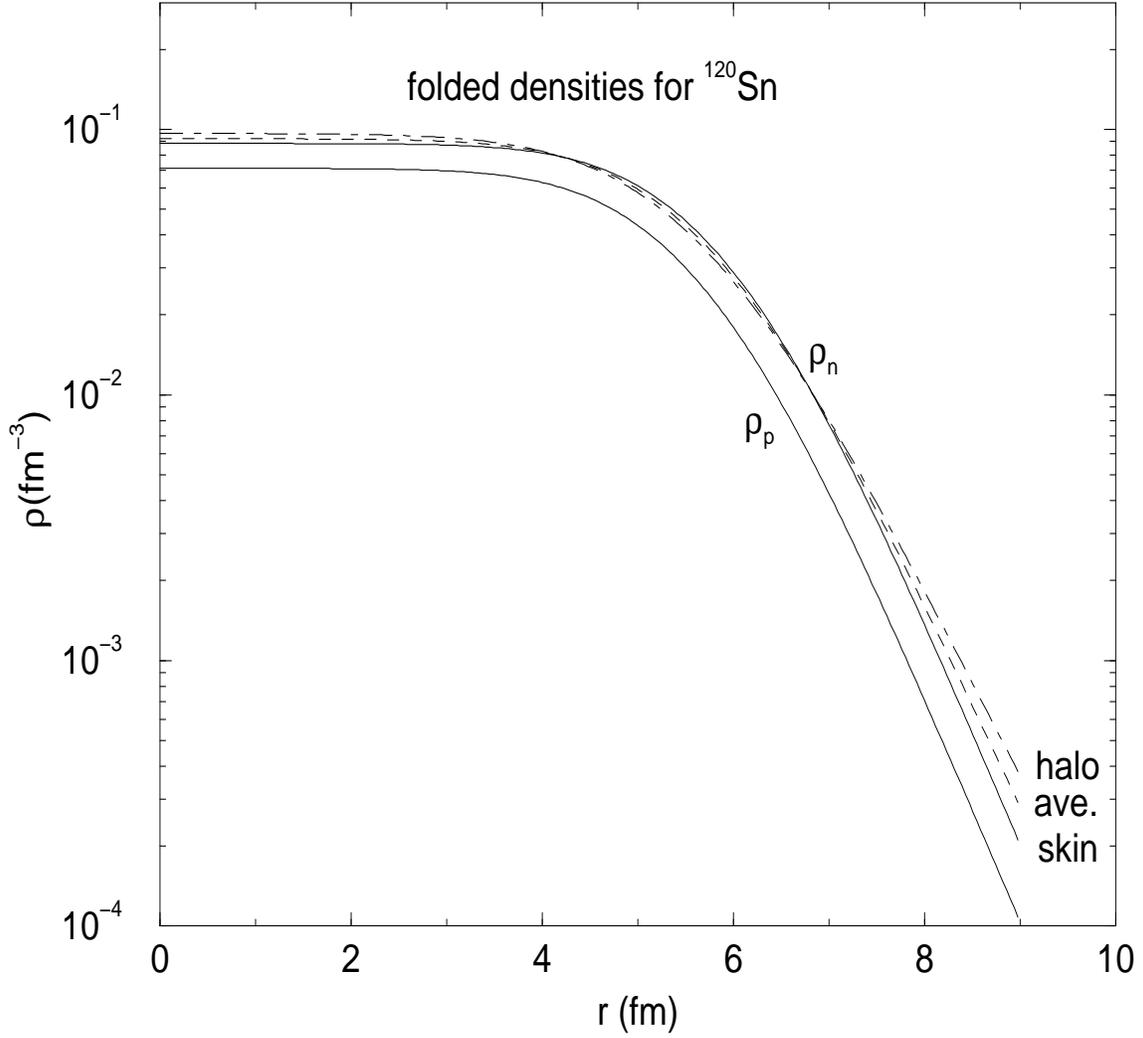, height=140mm,width=150mm}
\caption{Proton and neutron densities for $^{120}$Sn after a finite range
folding with $\beta$ =0.85 fm, see Eq.(\ref{equ:fold}). Neutron densities
are calculated with $\alpha$ = 1.2 fm, see Eq.(\ref{equ:RMF}).}
\label{fig0}
\end{figure}

\begin{figure}
\epsfig{file=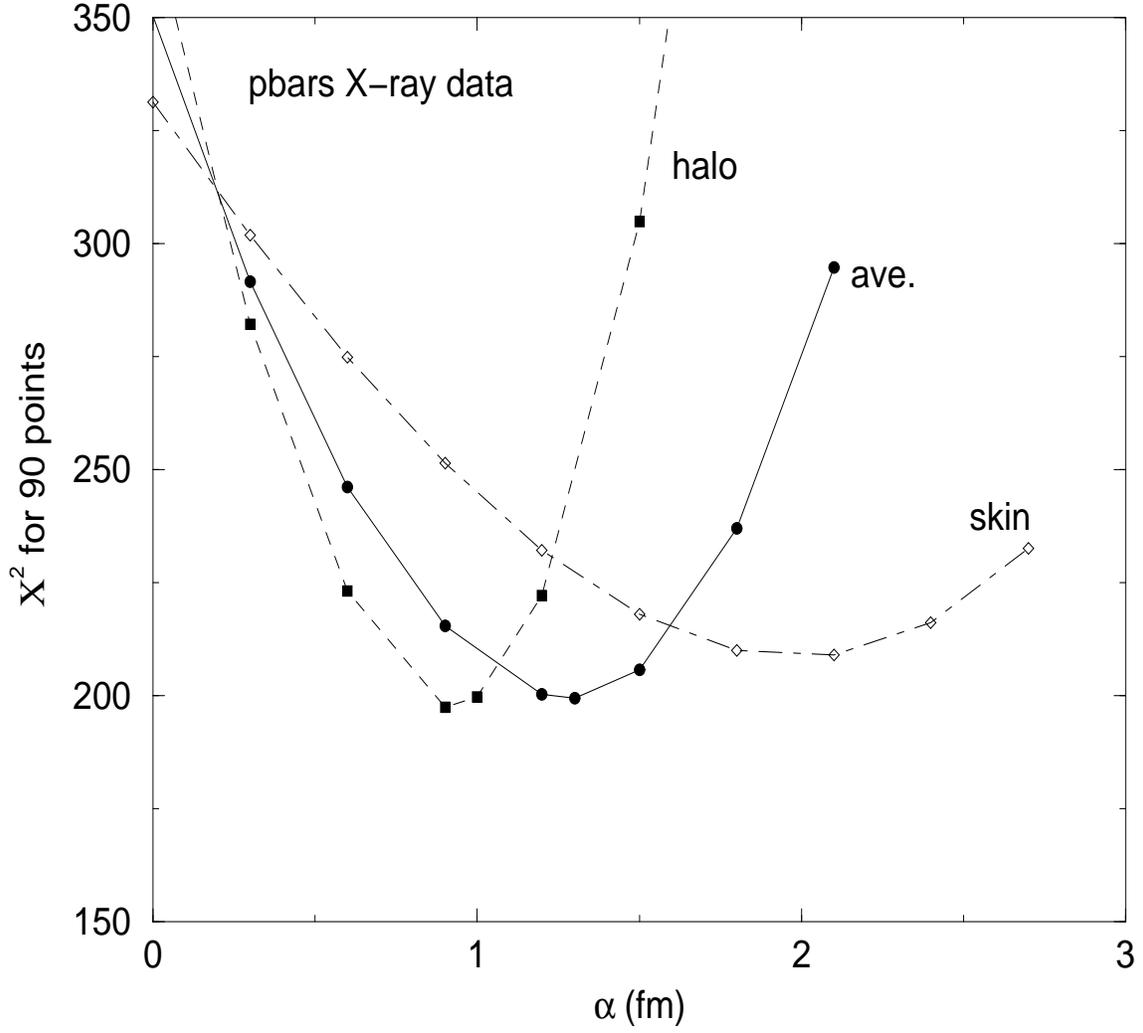, height=140mm,width=150mm}
\caption{$\chi ^2$ values from fits to $\bar p$ X-ray data as function
of the slope parameter of $r_n-r_p$ (Eq.(\ref{equ:RMF})) for various
types of neutron density.}
\label{fig1}
\end{figure}

\begin{figure}
\epsfig{file=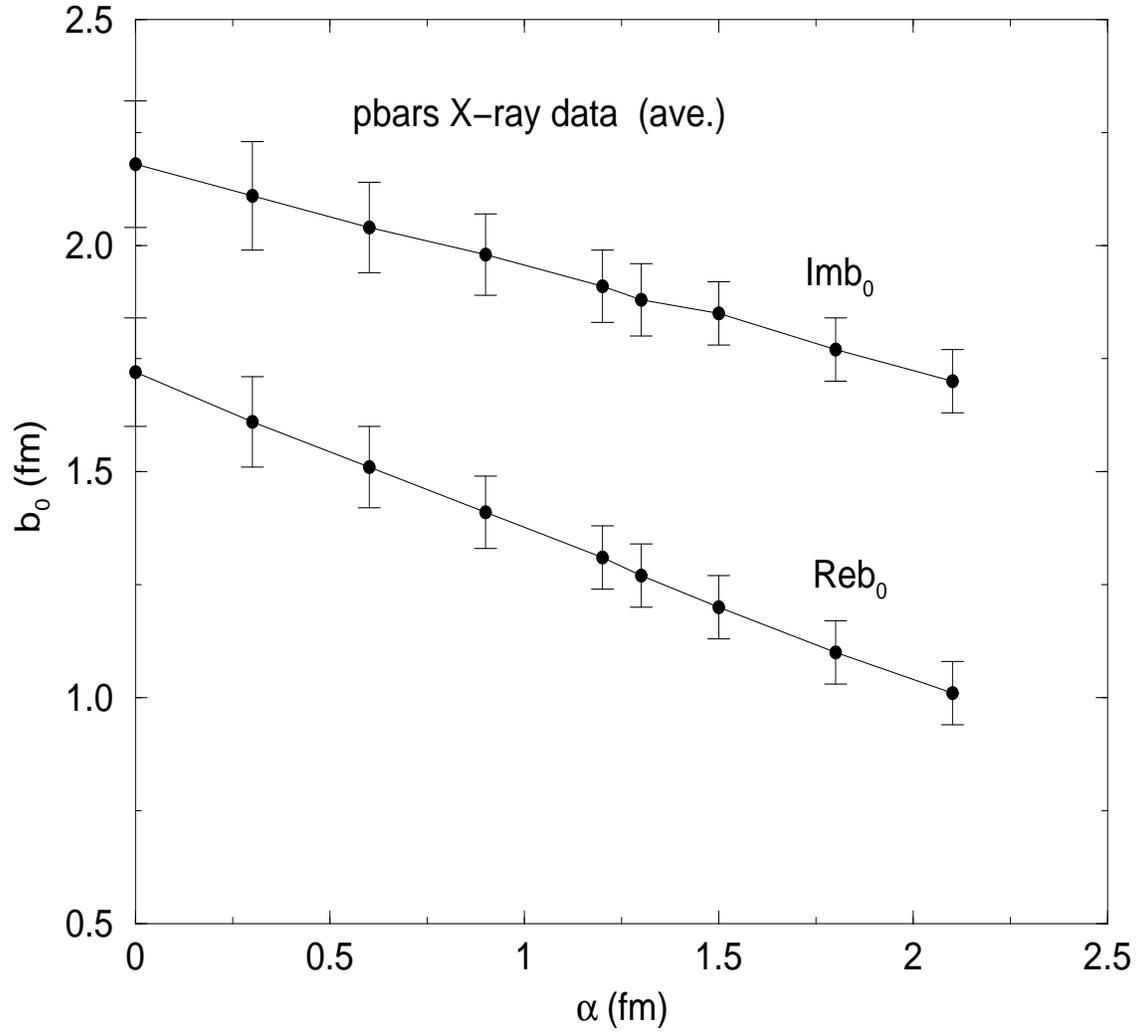, height=140mm,width=150mm}
\caption{Values of the potential parameter $b_0$ for the `average'
type of neutron
density.}
\label{fig2}
\end{figure}

\begin{figure}
\epsfig{file=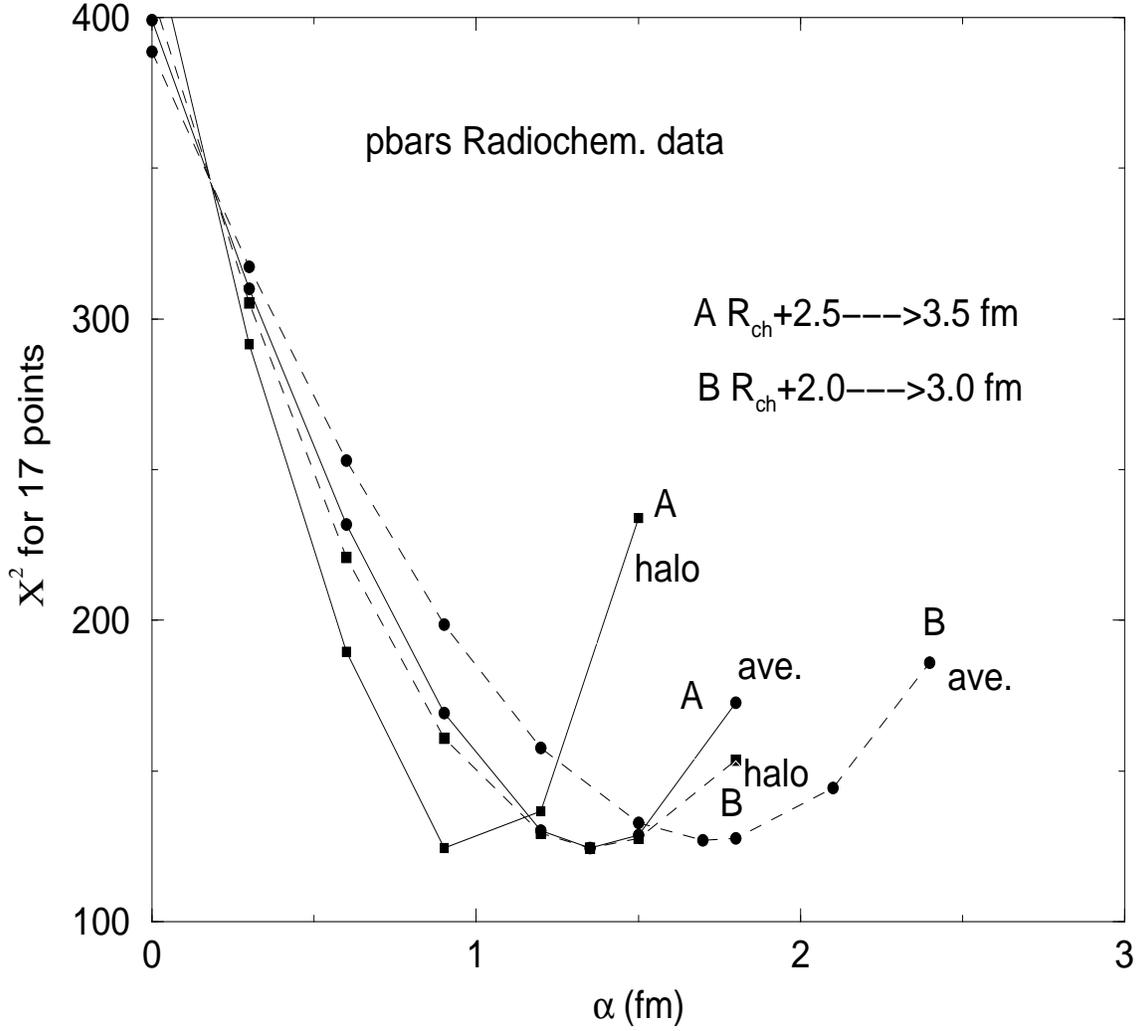, height=140mm,width=150mm}
\caption{$\chi ^2$ values from fits to radiochemical data as function
of the slope parameter of $r_n-r_p$ (Eq.(\ref{equ:RMF})) for `halo'
and `average' types of neutron density and for two integration ranges:
A for 2.5 to 3.5 fm beyond the charge radius; B for 2.0 to 3.0 beyond the
charge radius, see text.}
\label{fig3}
\end{figure}

\begin{figure}
\epsfig{file=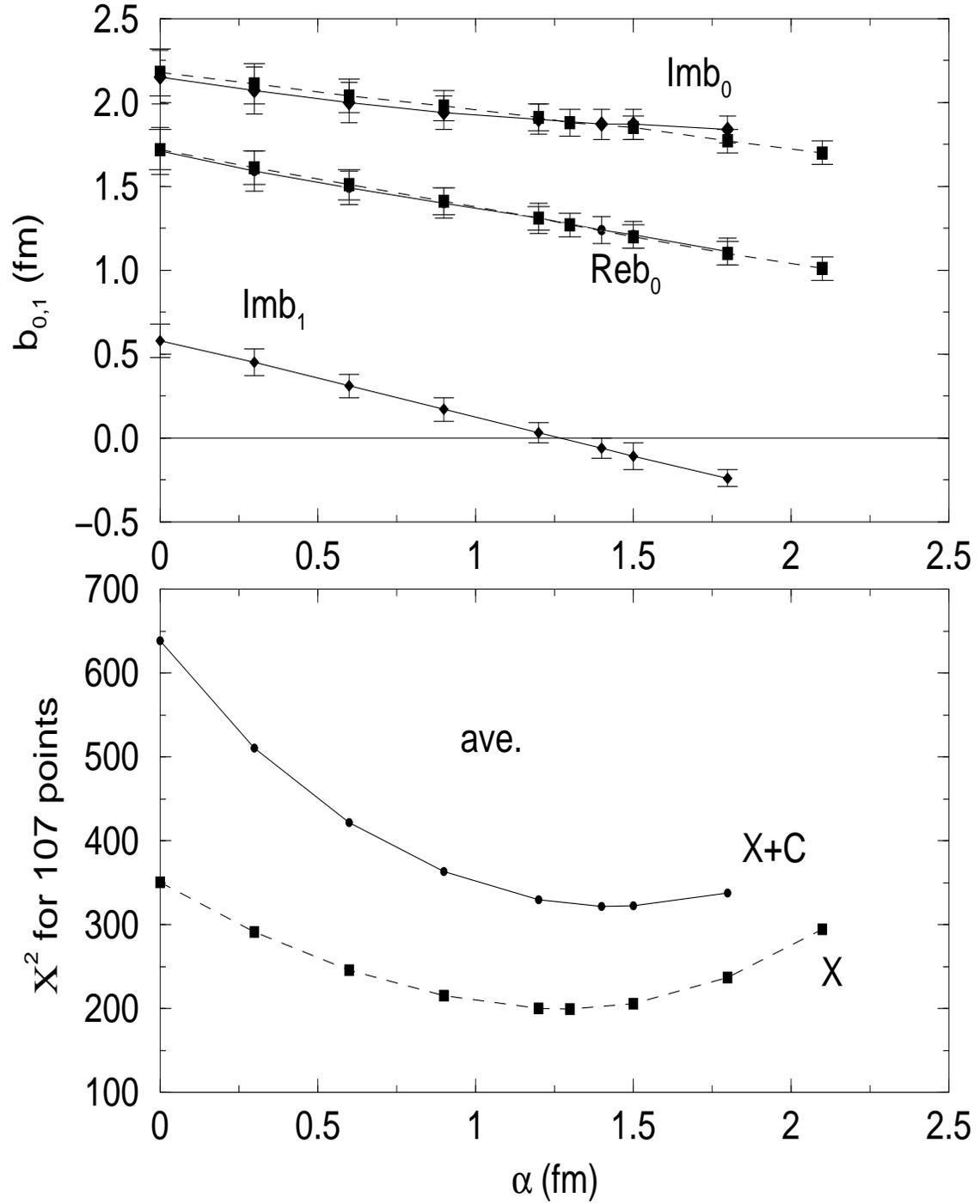, height=190mm,width=150mm}
\caption{Results of global fits for the `average' model of neutron densities.
Solid lines for combined X-ray and radiochemical data; dashed lines for
X-ray data only.}
\label{fig4}
\end{figure}

\begin{figure}
\epsfig{file=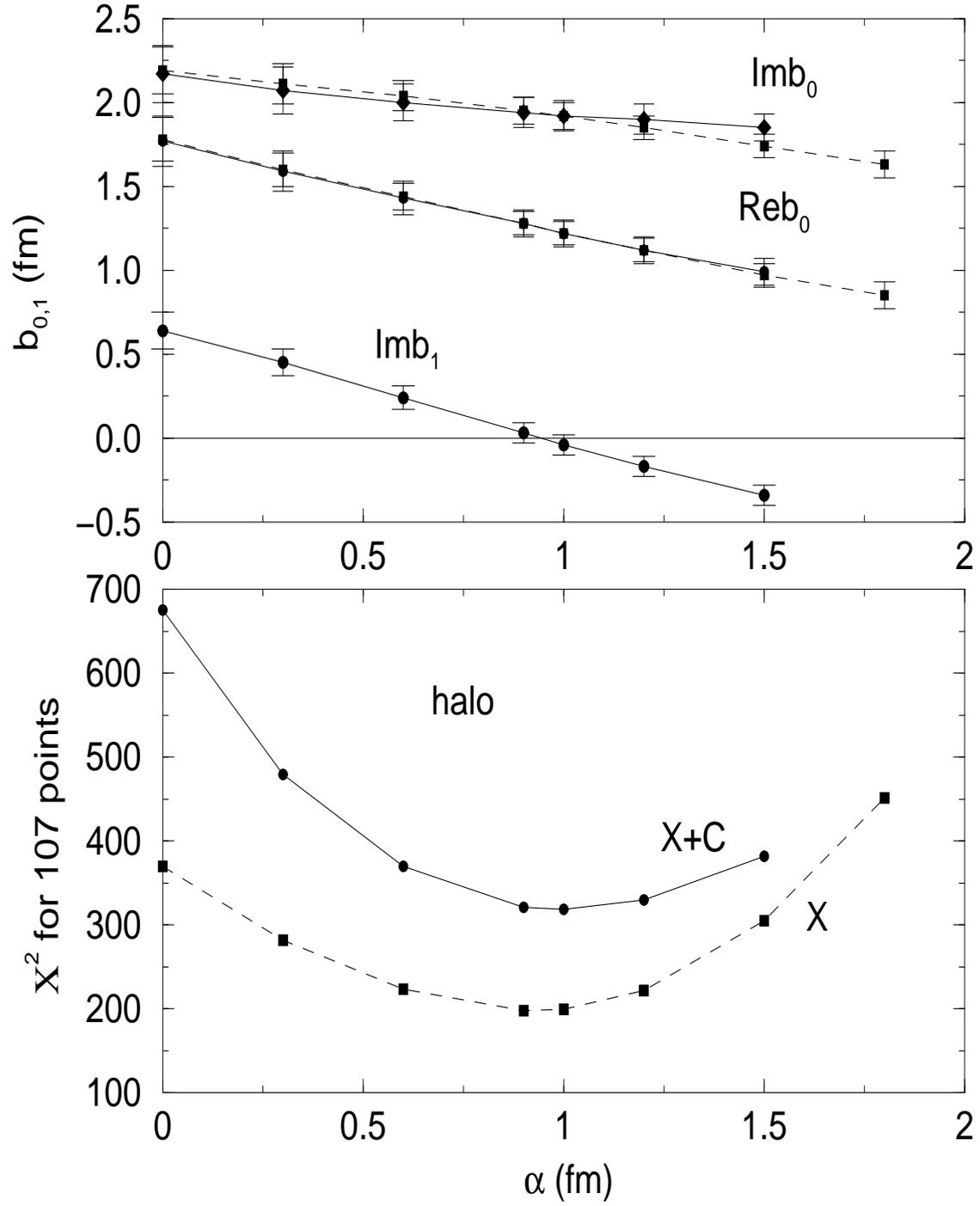, height=190mm,width=150mm}
\caption{Same as Fig. \ref{fig4} but for the `halo' type of neutron
densities.}
\label{fig5}
\end{figure}

\end{document}